\documentclass{sf2a-conf2011}
\usepackage{graphicx}
\usepackage{hyperref}
\usepackage[]{natbib}  
\usepackage[cyr]{aeguill}
\usepackage{epstopdf}

\def\BibTeX{{\rm B\kern-.05em{\sc i\kern-.025em b}\kern-.08em
    T\kern-.1667em\lower.7ex\hbox{E}\kern-.125emX}}
\bibpunct{(}{)}{;}{a}{}{,}  
\def\msun{\hbox{${\rm M}_{\odot}$}}
\def\mstar{\hbox{$M_{\star}$}}
\def\lstar{\hbox{$L_{\star}$}}
\def\lsun{\hbox{$L_{\odot}$}}
\def\rsun{\hbox{${\rm R}_{\odot}$}}
\def\rstar{\hbox{$R_{\star}$}}
\def\Prot{\hbox{$P_{\rm rot}$}}
\def\d{\hbox{$\rm d$}}
\def\kG{\hbox{$\rm kG$}}
\def\ie{\hbox{i.e. }}   
\def\eg{\hbox{e.g., }}  
\def\BI{\hbox{$B_{\rm I}$}}
\def\BV{\hbox{$B_{\rm V}$}}

\begin{document}

\TitreGlobal{SF2A 2011}


\title{Evidence for dynamo bistability among very low mass stars}

\runningtitle{Dynamo bistability in very low mass stars}

\author{J.~Morin}\address{Dublin Institute for Advanced Studies, School of
  Cosmic Physics, 31 Fitzwilliam Place, Dublin 2, Ireland}
\author{X.~Delfosse}\address{Universit\'e. J.~Fourier (Grenoble 1)/CNRS;
  Laboratoire d'Astrophysique de Grenoble (LAOG, UMR 5571); F--38041 Grenoble,
  France}
\author{J.-F.~Donati}\address{IRAP-UMR 5277, CNRS \& Univ de Toulouse, 14 av E
  Belin, F-31400}
\author{E.~Dormy}\address{MAG (ENS/IPGP), LRA, Ecole Normale Sup\'erieure, 24
  Rue Lhomond, 75252 Paris Cedex 05, France}
\author{T.~Forveille$^2$}
\author{M.M.~Jardine}\address{School of Physics and Astronomy, Univ.\ of
  St~Andrews, St~Andrews, Scotland KY16 9SS, UK}
\author{P.~Petit$^3$}
\author{M.~Schrinner$^4$}

\setcounter{page}{237}

\index{Morin, J.}
\index{Delfosse, X.}
\index{Donati, J.-F.}
\index{Dormy, E.}
\index{Forveille, T.}
\index{Jardine, M.M.}
\index{Petit, P.}
\index{Schrinner, M.}


\maketitle


\begin{abstract}
Dynamo action in fully convective stars is a debated issue that also questions
our understanding of magnetic field generation in partly convective Sun-like
stars. During the past few years, spectropolarimetric observations have
demonstrated that fully convective objects are able to trigger strong
large-scale and long-lived magnetic fields. We present here the first
spectropolarimetric study of a sample of active late M dwarfs (M5-M8) carried
out with ESPaDOnS@CFHT. It reveals the co-existence of two distinct types of
magnetism among stars having similar masses and rotation rates. A possible
explanation for this unexpected discovery is the existence of two dynamo
branches in this parameter regime, we discuss here the possible identification
with the weak \textit{vs} strong field bistability predicted for the geodynamo.
\end{abstract}



\section{Introduction}
In  cool stars, which possess a convective envelope, magnetism is thought
to be constantly regenerated against ohmic decay by dynamo effect.  For
Sun-like stars the interface layer between the inner radiative zone and the
outer convective envelope is generally thought to play a major role in the
dynamo processes \cite[see \eg][]{Charbonneau10}. Since fully-convective stars
-- either main sequence stars below 0.35~\msun\ (\ie with spectral type later
than $\sim$ M4) or young T Tauri stars -- do not possess such an interface
layer, generation of magnetic field in their interiors is often thought to rely
on a non-solar-type dynamo. However, the precise mechanism and the properties
of the resulting magnetic have been a debated issue
\cite[][]{Durney93, Chabrier06, Dobler06, Browning08}. 
 
Two main complementary approaches are successfully applied to study magnetic
fields close to the fully-convective transition. On the one hand, by modelling
Zeeman broadening of photospheric spectral lines it is possible to assess the
magnetic field averaged over the visible stellar disc \cite[\eg][]{Reiners06}.
This method is therefore able to probe magnetic fields regardless of their
complexity but provides very little information about the field geometry. On
the other hand, the Zeeman-Doppler imaging technique models the evolution of
polarization in spectral lines during at least one rotation period in order to
reconstruct a map of the large-scale component of the vector magnetic field on
the stellar photosphere.

Spectropolarimetric studies of a sample of M0--M4 dwarfs, conducted
with ESPaDOnS and NARVAL, have revealed for the first time a
strong change in large-scale magnetic topologies occurring close to the
fully-convective boundary. Stars more massive than 0.5~\msun\ exhibit
large-scale fields of moderate intensity featuring a significant toroidal
component and a strongly non-axisymmetric poloidal component, with evolution
happening on a timescale of less than 1~yr \cite[][D08]{Donati08b}. For those
in the range 0.25--0.50~\msun\ much stronger large-scale fields are observed,
which are dominated by the axial dipolar component and show only very limited
evolution over successive years \cite[][M08a,b]{Morin08a, Morin08b}.
Comparisons of these large-scale magnetic field measurements with X-ray
activity indices or with measurements of the total magnetic field (\ie at all
spatial scales) derived from the analysis of Zeeman broadening of FeH molecular
lines, suggest that fully-convective stars are much more efficient at
generating large-scale magnetic field than partly-convective ones
\cite[D08,][]{Reiners09b}.

\section{Surface magnetic fields of late M dwarfs}

\begin{figure}[ht!]
 \centering
 \includegraphics[width=0.8\textwidth,clip]{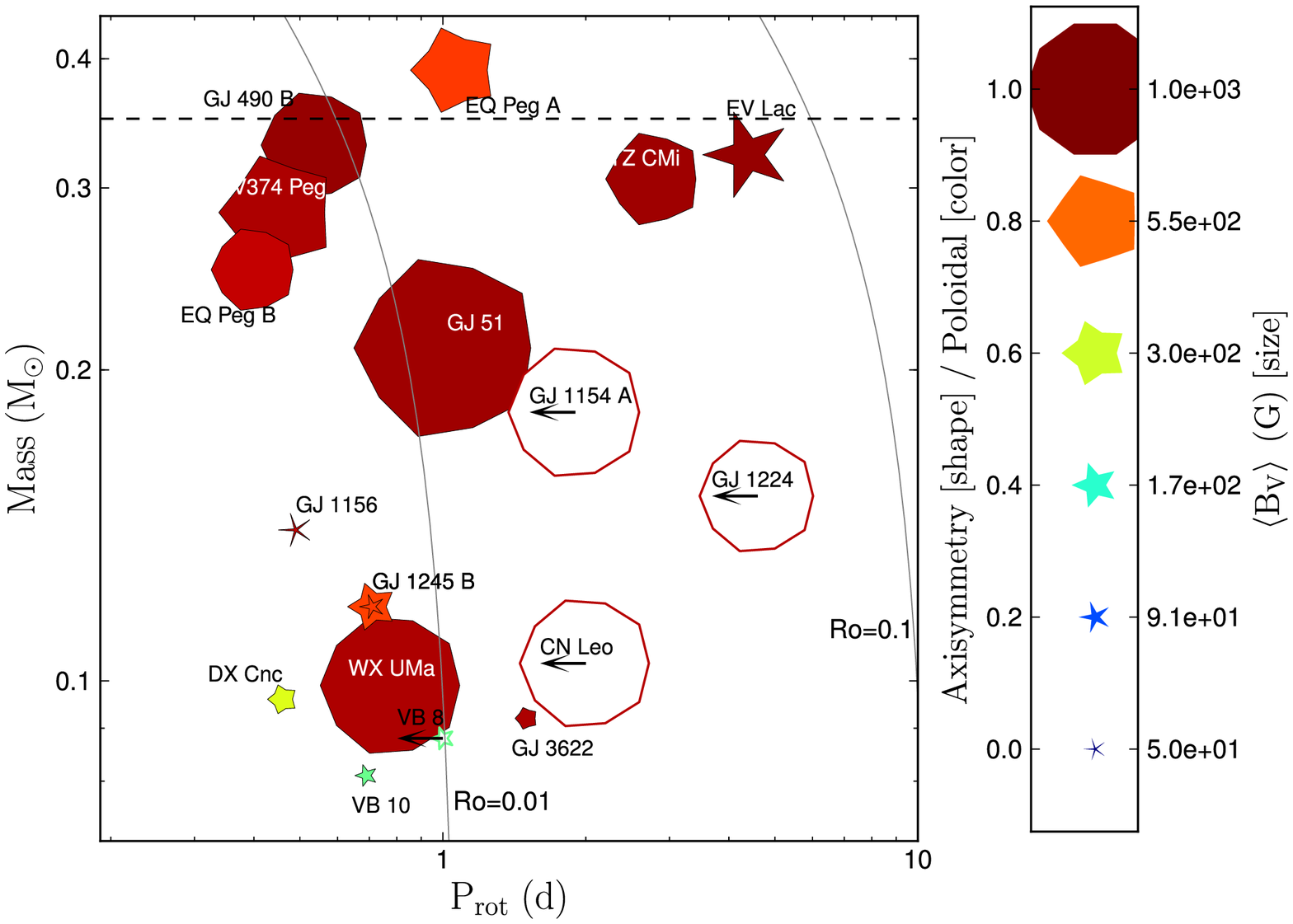}
  \caption{Mass--period diagram of fully-convective stars derived from
spectropolarimetric data and Zeeman-Doppler Imaging (ZDI). Symbol size
represents the reconstructed magnetic energy, the color ranges from blue for a
purely toroidal to red for a purely poloidal field, and the shape depicts the
degree of axisymmetry from a sharp star for non-axisymmetric to a regular
decagon for axisymmetric. For a few stars of the sample \cite{Morin10a} could
not perform a definite ZDI reconstruction,  in these cases only an upper limit
of the rotation period is known and the magnetic flux is extrapolated, those
objects are depicted as empty symbols.  The theoretical fully-convective limit
is depicted as a horizontal dashed line. Thin solid lines represent contours of
constant Rossby number Ro=0.01 (left) and 0.1 (right), as estimated in
\cite{Morin10a}.}
  \label{morin:fig1}
\end{figure}

\begin{figure}[ht!]
 \centering
 \includegraphics[width=0.485\textwidth,clip]{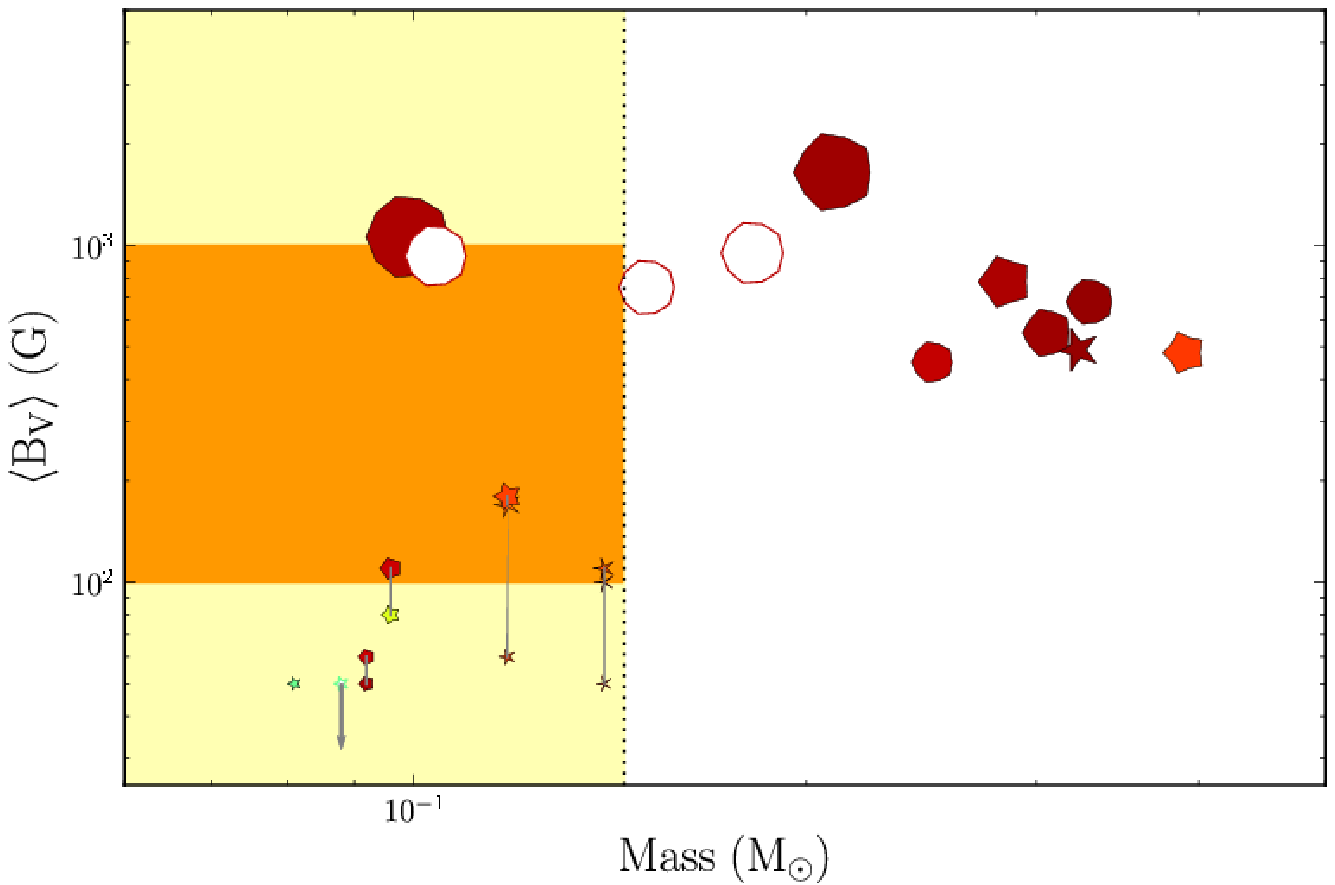}\hfill%
 \includegraphics[width=0.485\textwidth,clip]{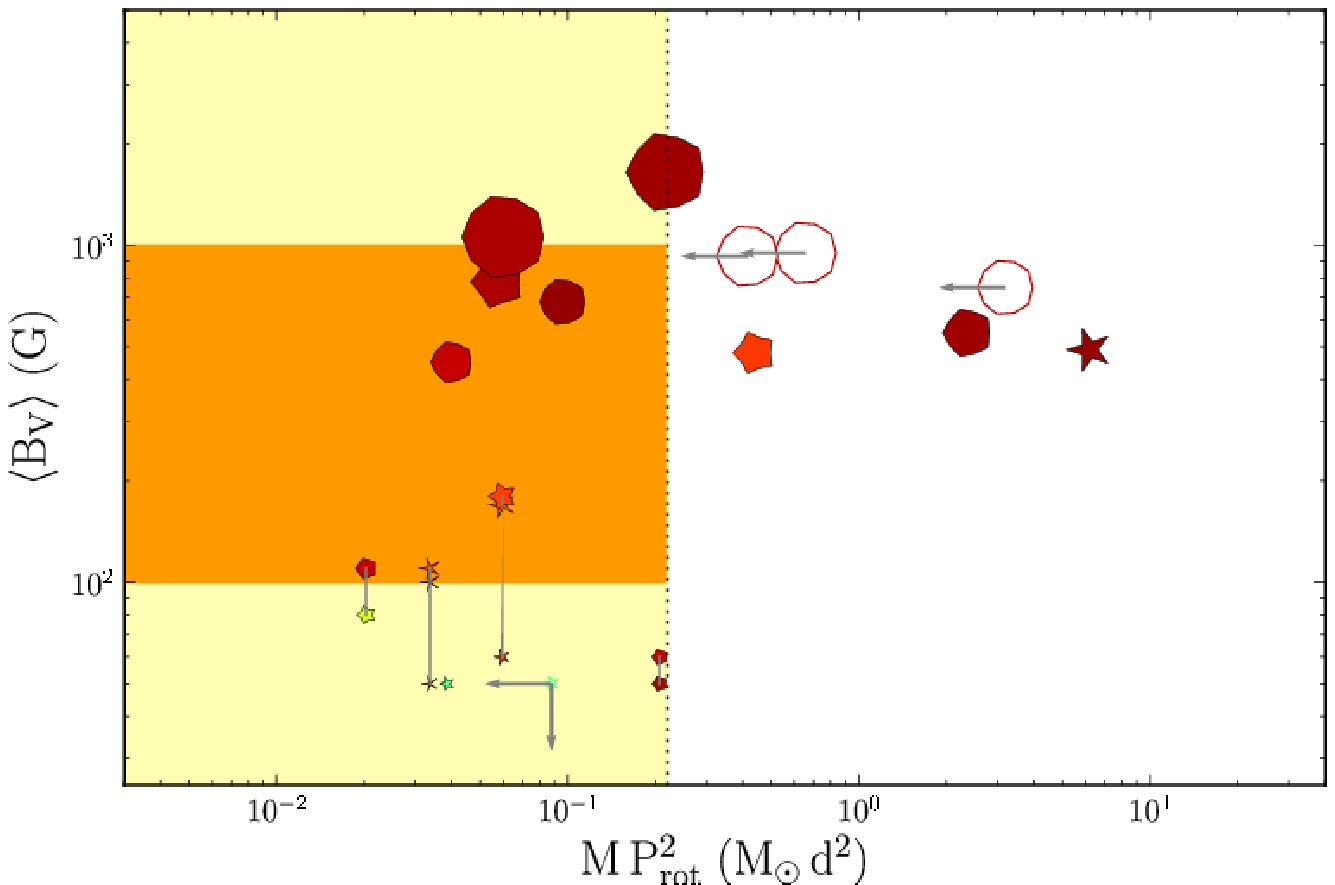}\hfil\\
  \caption[]{Average large-scale magnetic fluxes of fully-convective stars
derived from
spectropolarimetric data and Zeeman-Doppler Imaging (ZDI), as a function of
mass ({\bfseries left panel}) and mass $\times\,\Prot^2$ ({\bfseries right
panel}). Symbols are similar
to those used in the mass--period diagram (see Fig.~\ref{morin:fig1}). For stars
in
the WM regime symbols corresponding to different epochs for a given star are
connected by a vertical grey line. The yellow region represents the domain
where bistability is observed and the orange one separates the two types of
magnetic fields identified (see text).}
  \label{morin:fig23}
\end{figure}

A sample of 11 active M dwarfs with masses significantly below the
fully-convective boundary ($0.08<\mstar<0.21~\msun$ or spectral types M5--M8)
has been observed with the ESPaDOnS spectropolarimeter \cite[][hereafter
M10]{Morin10a}. Below 0.15~\msun, we observe two radically different categories
of large-scale magnetic fields: either a strong and steady almost dipolar field
(hereafter SD, similar to stars in the range 0.15--0.5~\msun); or a weaker
multipolar, non-axisymmetric field configuration undergoing dramatic evolution
on a timescale of at most 1~yr (hereafter WM). However the two groups of
objects cannot be separated in a mass-rotation diagram, see
Fig.~\ref{morin:fig1}. No object is observed to evolve from one type of
magnetism to the other during the survey (some objects were observed for 4
years). In terms of large-scale magnetic field values, a gap exists between
these two types of magnetism, with no object with $200<\BV<900~G$ in this mass
range, see Fig.~\ref{morin:fig45}. Both stars hosting weak multipolar (WM) or
strong dipolar (SD) fields have very strong total magnetic fields (2--4~\kG).
No systematic correlation is found between the type of large-scale magnetic
topology and the total magnetic field \BI\ (see Fig.~\ref{morin:fig23}). Hence,
the two different types of magnetic field configurations are only detected when
considering the large-scale component (probed by spectropolarimetry, and which
represents 15-30~\% of the total flux in the SD regime, but only a few percent
in the WM regime) and not the total magnetic flux derived from unpolarised
spectroscopy.  This unexpected observation may be explained in several
different ways: for instance, another parameter than mass and rotation period
(such as age) may be relevant, two dynamo modes may be possible or stars may
switch between two states in this mass range, etc.

\section{Weak and strong field dynamos: from the Earth to the stars}

\begin{figure}[ht!]
 \centering
 \includegraphics[width=0.485\textwidth,clip]{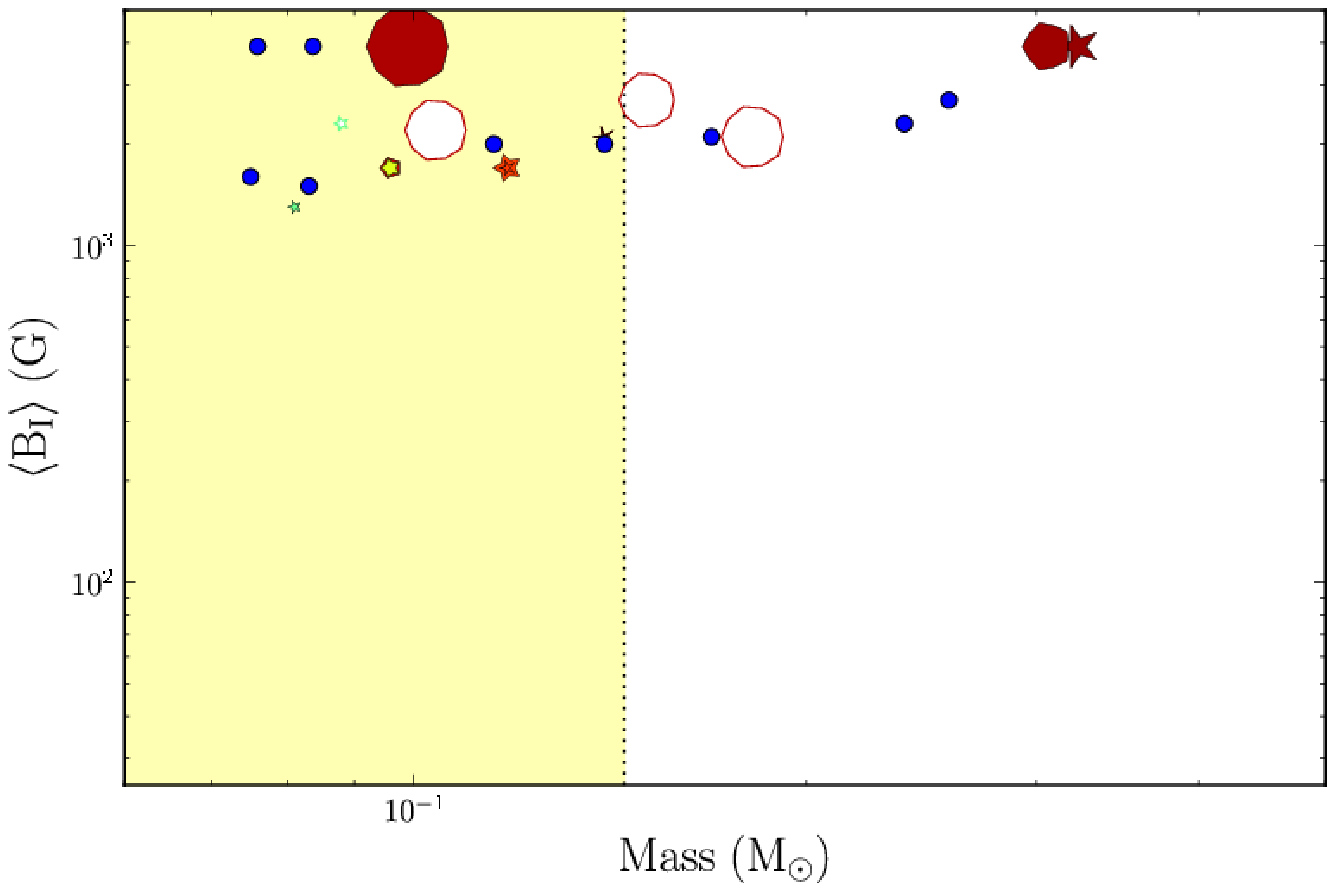}\hfill%
 \includegraphics[width=0.40\textwidth,clip]{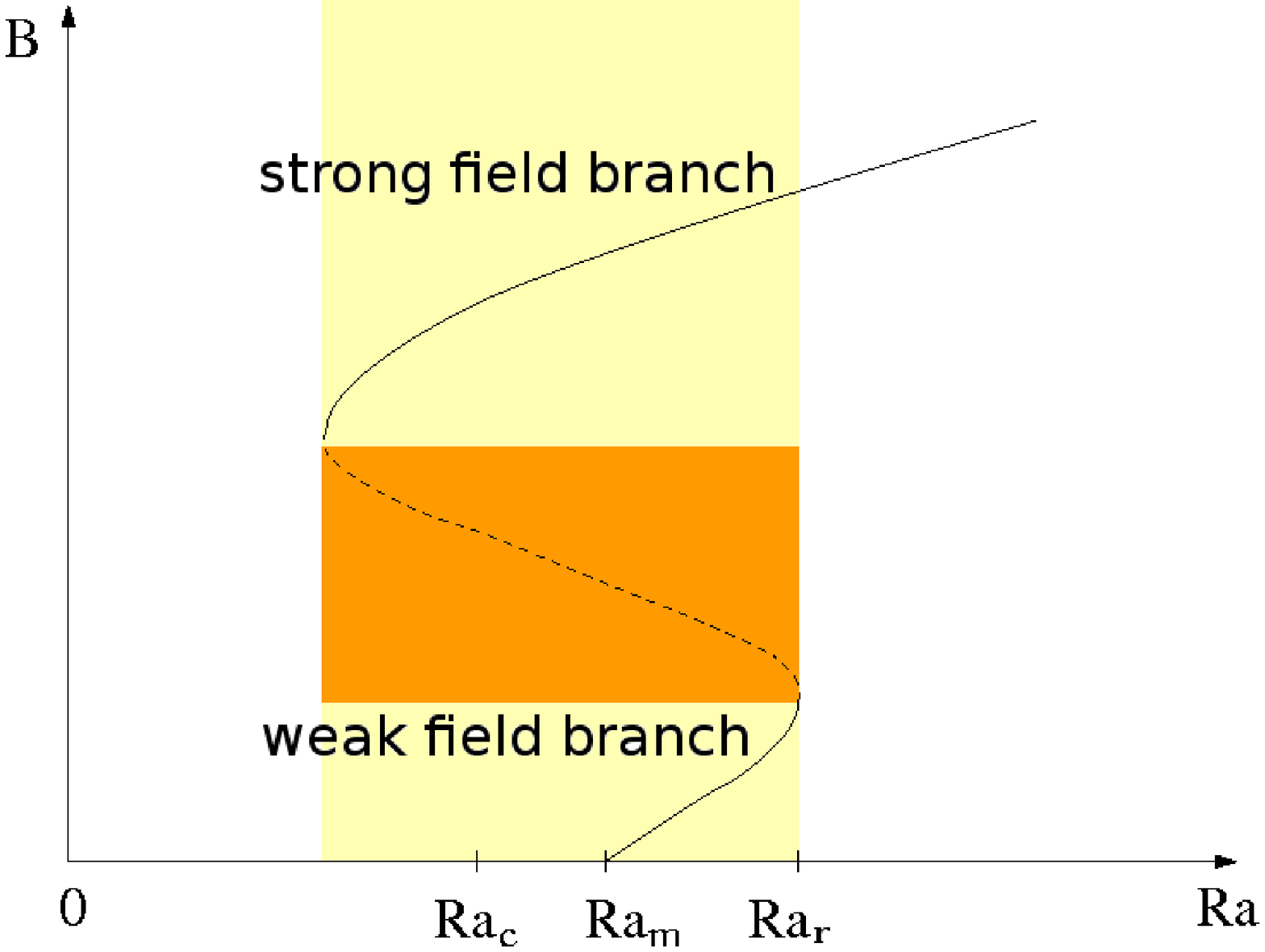}\hfil\\
  \caption{{\bfseries Left:} Total magnetic fluxes of fully-convective
stars measured from unpolarised spectra of FeH lines. The values are taken from
\cite{Reiners09} and \cite{Reiners10}, whenever 2MASS near infrared
luminosities \cite[][]{Cutri03} and Hipparcos parallaxes \cite[][]{ESA97} are
available to compute the stellar mass from the \cite{Delfosse00}
mass--luminosity relation.  Whenever spectropolarimetric data are available the
properties of the magnetic topology are represented as symbols described in
Fig.~\ref{morin:fig23}. Else small blue symbols are used, upward (downward)
triangles represent lower (upper) limits.  {\bfseries Right:} Anticipated
bifurcation diagram for the geodynamo \cite[adapted from][]{Roberts88}. The
magnetic field amplitude is plotted against the Rayleigh number. The
bifurcation sequence is characterised by two branches, referred to as weak and
strong field branches. The yellow and orange regions have the same meaning as
in Fig.~\ref{morin:fig23}. $\rm Ra_c$ is the critical Rayleigh number for the
onset of non-magnetic convection. The weak field regime sets in at $ \rm Ra_m$,
and the turning point associated with the runaway growth corresponds to ${\rm
Ra} = {\rm Ra_r}$.} \label{morin:fig45} \end{figure}

In this section we briefly detail one of the hypothesis that could explain the
observation of two groups of late M dwarfs with very different magnetic
properties: the fact that two different dynamo modes could genuinely operate in
stars having very similar mass and rotation. We focus here on the weak
\textit{vs} strong field dynamo bistability, initially proposed for the
geodynamo. The underlying idea is that both, magnetic fields and rotation,
taken separately tend to inhibit convection, but that if both effects are
combined the impeding influences of the Lorentz and of the Coriolis forces may
be reduced, allowing convection to set on at lower Rayleigh number and to
develop on larger length scales \cite[][]{Chandra61}. This led \cite{Roberts78}
to conjecture that in a rapidly rotating system, for magnetic fields stronger
than a threshold value, the Lorentz force would enhance convection and hence
dynamo action, resulting in a runaway growth of the magnetic field. The
corresponding bifurcation diagram is depicted on Fig.~\ref{morin:fig45}. On the
weak-field branch the Lorentz force is balanced by viscous or inertial terms in
the momentum equation, this force balance requires small-spatial scales. On the
strong field branch, however, the magnetic field strength is set by a balance
between Lorentz and Coriolis forces, which requires larger spatial scales, this
is the magnetostrophic regime. A similar bifurcation diagram, but based on the
fact that magnetic buoyancy would be negligible close to the dynamo onset has
been proposed for stars by \cite{WT2000}.

We now briefly discuss the identification between WM (SD) magnetism and
weak-field (strong-field) dynamo regime, the reader is referred to
\cite{Morin11} for a more detailed discussion. The usual control parameter in
the weak vs strong field dynamo scenario described above is the Rayleigh
number, which measures the energy input relative to forces opposing the motion.
Mass can be used as a good proxy for the available energy flux in M dwarfs,
Fig.~\ref{morin:fig23} can therefore be interpreted as a bifurcation diagram
for the amplitude of the large scale magnetic field versus a control parameter
measuring the energy input. In order to compare the driving of convection with
the impeding effect of rotation, we can use $\mstar\times\Prot^2$ as a rough
proxy for the Rayleigh number (see Fig.~\ref{morin:fig23}) based on rotation
rather than diffusivities \cite[\eg][]{Christensen06}.

First, in the SD regime the magnetic field strength has to be compatible with a
Lorentz--Coriolis force balance. We note that this balance is valid spatial
large-scales for which the Coriolis term is predominant over inertial terms in
the momentum equation, in qualitative agreement with the observation that only
the large-scale component of the magnetic field exhibits a bimodal
distribution. This magnetostrophic force balance roughly corresponds to an
Elsasser number of order unity, \ie:
\begin{equation} \Lambda = \frac{B^2}{\rho\mu\eta\Omega} \sim 1, \end{equation}
where $B$ is the magnetic field strength, $\rho$ the mass density, $\mu$ the
magnetic permeability, $\eta $ the magnetic diffusivity and $\Omega$ the
rotation rate. With a few assumptions described in \cite{Morin11}, we find that
the order of magnitude of the expected magnetic field strength on the strong
field branch is set by:
\begin{equation} \small {B_{sf}} \sim 6\,
\left(\frac{\mstar}{\msun}\right)^{1/2} %
\left(\frac{\rstar}{\rsun}\right)^{-1}  \left(\frac{\lstar}{\lsun}\right)^{1/6}
\left(\frac{\eta_\odot}{\eta_{\rm ref}}\right)^{1/2} %
\left(\frac{\Prot}{1~\d}\right)^{{-}1/2}~\kG \end{equation}
Where $\eta_\odot$ is a reference value for the magnetic diffusivity in the
solar convection zone, and $\eta_{\rm ref} = 10^{11}~{\rm cm^2\,s^{-1}}$.
Taking stellar radius and luminosity for the stellar mass in the range
$0.08-0.35~\msun$ from \cite{Chabrier97} main sequence models, and $\eta_\odot$
in the range $10^{11}-3\times10^{12}~{\rm cm^2\,s^{-1}}$
\cite[\eg][]{Ruediger11}, we derive surface values in the strong field regime
in the range 2--50~\kG, compatible with the order of magnitude of measured \BV\
values. More conclusively, the gap in terms of \BV\ between the two branches
depends on the ratio of inertia to Coriolis force in the momentum equation and
can estimated with:
\begin{equation} \frac{B_{wf}}{B_{sf}} = Ro^{1/2}, \end{equation}
which is of the order of $10^{-1}$ for stars of our sample in the bistable
domain, in good agreement with the typical ratio of large-scale magnetic fields
measured between the WM and SD groups of stars (see Fig.~\ref{morin:fig23}).

We note that according to the \cite{Chabrier97} main sequence models, the
product of the terms depending on stellar mass, radius and luminosity in the
expression of $B_{sf}$ is almost constant in the mid-to-late M dwarf regime.
The expected magnetic field strength on the strong field branch hence almost
scales with $\Omega^{1/2}$. This is not in contradiction with the fact all the
stars in our sample belong to the so-called saturated regime of of the
rotation--activity relation. Indeed $B_{sf} \propto \Omega^{1/2}$ (derived from
$\Lambda \sim 1$) should apply here to the large scale field alone, which is
only a fraction of the total magnetic field of the stars (between 15 and 30
\%). If a small scale dynamo operates, it does not need to follow the same
dependency. Finally, the weak dependency of the large-scale magnetic field on
stellar rotation predicted for stars in the strong-field regime cannot be ruled
out by existing data and should be further investigated.

\section{Conclusions}
We present here the main results of the first spectropolarimetric analysis of a
sample of active late M dwarfs \cite[more throughly detailed in][]{Morin10a}.
In particular we report the co-existence of two radically different types of
magnetism -- strong and steady dipolar field (SD) as opposed to weaker
multipolar field evolving in time (WM) -- for stars with very similar masses
and rotation periods. One of the foreseen hypothesis to explain these
observations is the genuine existence of two types of dynamo in this parameter
regime, \ie bistability. We show that the weak \textit{vs} strong field dynamo
bistability is a promising frame work. The order of magnitude of the observed
magnetic field in stars hosting a strong dipolar field, and more conclusively
the typical ratio of large-scale magnetic fields measured in the WM and SD
groups of stars are compatible with theoretical expectations. We argue that the
weak dependency of the magnetic field on stellar rotation predicted for stars
in the strong-field regime cannot be ruled out by existing data and should be
further investigated.  We do not make any prediction on the extent of the
bistable domain in terms of stellar parameters mass and rotation period, this
issue shall be investigated by further theoretical work, and by surveys of
activity and magnetism in the ultracool dwarf regime.

A dynamo bistability
offers the possibility of hysteretic behaviour. Hence the magnetic properties
of a given object depend not only on its present stellar parameters but also on
their past evolution.  For instance, for young objects episodes of strong
accretion can significantly modify their structure and hence the convective
energy available to sustain dynamo action \cite[][]{Baraffe09} initial
differences in rotation periods of young stars could also play a role. Because
stellar magnetic fields are central in most physical processes that control the
evolution of mass and rotation of young stars \cite[in particular
accretion-ejections processes and star-disc coupling,
\eg][]{Bouvier09,Gregory10}, the confirmation of stellar dynamo bistability
could have a huge impact on our understanding of formation and evolution of low
mass stars.




%
\end{document}